\documentclass[aps,prapplied,reprint,twocolumn]{revtex4-2}
\usepackage{amssymb}
\usepackage{graphicx}
\usepackage{amsmath}
\usepackage{epsfig}
\usepackage[latin9]{inputenc}
\usepackage{array}
\usepackage{multirow}
\usepackage{color}
\usepackage{esint}
\usepackage{bm}
\usepackage{textcomp}
\usepackage{bbm}
\usepackage{epstopdf}
\usepackage{mathtools}
\usepackage{soul}
\usepackage[dvipsnames]{xcolor}
\usepackage{tikz}
\usepackage{tabularx}
\usepackage{float}
\usepackage[bookmarks=true,pdftex,colorlinks=true,urlcolor=blue,linkcolor=black,citecolor=blue,breaklinks=true,hypertexnames=false]{hyperref}
\usepackage{multibib} 
\usepackage{physics}
\usepackage{makecell}
\usepackage{booktabs}

\begin{document}

\title{Optical transport of cold atoms to quantum degeneracy}

\author{Yanqing Tao}
\affiliation{State Key Laboratory for Mesoscopic Physics and Frontiers Science Center for Nano-optoelectronics, School of Physics, Peking University, Beijing 100871, China}

\author{Yufei Wang}
\affiliation{State Key Laboratory for Mesoscopic Physics and Frontiers Science Center for Nano-optoelectronics, School of Physics, Peking University, Beijing 100871, China}

\author{Ligeng Yu}
\affiliation{State Key Laboratory for Mesoscopic Physics and Frontiers Science Center for Nano-optoelectronics, School of Physics, Peking University, Beijing 100871, China}

\author{Bo Song}\email{bsong@pku.edu.cn}
\affiliation{State Key Laboratory for Mesoscopic Physics and Frontiers Science Center for Nano-optoelectronics, School of Physics, Peking University, Beijing 100871, China}

\begin{abstract}

Efficient transport of cold atoms is essential for continuous operation, enabling applications ranging from atomic lasers to continuously operated qubits. However, deep potentials required to overcome vibrations, axial trap nonuniformity and insufficient cooling have limited transport of cold atoms near quantum degeneracy. Here we demonstrate rapid optical transport of cold atoms to Bose-Einstein condensation using a moving optical lattice formed by two Bessel beams. A gas of $3 \times 10^5$ ytterbium atoms at a temperature of $340\,$nK is transported over $34\,$cm in $350\,$ms with efficiency over $60\%$. Furthermore, a degenerate gas of $1 \times 10^5$ atoms with a $40\%$ condensate fraction emerges from the phase synchronization process driven by atomic interactions. This demonstration enables the fast preparation of ultracold atomic beams and large-scale atom arrays for quantum sensing, simulation and computing.

\end{abstract}

\maketitle

\section{Introduction}

Atom transport has been widely employed to deliver atoms into better vacuum conditions for longer lifetimes. It also plays a crucial role in the preparation and continuous loading of cold atoms for applications ranging from atom lasers in quantum sensing to atomic qubits in quantum simulation and computing~\cite{chen2022continuous,chiu2025continuous,tao2024high,schafer2025continuous,yu2026conveyor}. Although transport has been experimentally demonstrated using different methods including magnetic~\cite{sauer2004cavity,nakagawa2005simple,pertot2009versatile,Handel2011Magnetic,greiner2001magnetic,hansel2001magnetic,minniberger2014magnetic}, optical~\cite{gustavson2002transport,couvert2008optimal,naides2013trapping,leonard2014optical,schrader2001applied,schmid2006long,klostermann2022fast,bao2022fast,trisnadi2022design,matthies2024long}, and hybrid schemes~\cite{pritchard2006transport,marchant2011guided}, fast transport of cold atoms near degeneracy over long distances remains challenging, due to vibrations, nonuniform potentials along the transport and insufficient cooling.

In this letter, we demonstrate the transport of cold atoms using a moving optical lattice formed by two Bessel beams with precise control. Owing to the properties of Bessel beams, their extended nondiffracting range along the transport direction exceeds that of Gaussian beams of the same beam size, allowing for more uniform and shallower potentials for transporting colder atoms. We manage to transport a gas of \(3 \times 10^5\) ytterbium (Yb) atoms at $340\,$nK over $34\,$cm in $350\,$ms, with position control accuracy of $2\,\mu$m over the whole transport.

Eventually a degenerate gas of $1\times10^5$ atoms with a $40\%$ condensate fraction is produced through collision thermalization followed by a phase synchronization process. By both decelerating the moving lattice and lowering the trap potential in the final stage, the hotter atoms are spilled out from each pancake trap in the moving lattice. Notably, the deceleration effectively lowers the trap depth, but the trap frequencies remain high, maintaining a high atomic density favorable for evaporative cooling. The moving lattice separates approximately 57 pancake-shaped clouds, each with a random phase resulting from independent collisional thermalization. After releasing the atoms into a shallow dipole trap, the phases of those pancakes are synchronized by atomic interactions within a few hundred milliseconds, leading to Bose-Einstein condensation.

Our method not only enables rapid optical transport of cold atoms but also facilitates the fast realization of degenerate gases~\cite{hu2017creation,urvoy2019direct,xin2025fast}, and is broadly applicable to atom lasers and continuously operated large-scale atomic arrays~\cite{chen2022continuous,chiu2025continuous,tao2024high}.

\section{Experimental setup}

\begin{figure}[htbp]
    \centering
    \includegraphics[width=1\linewidth]{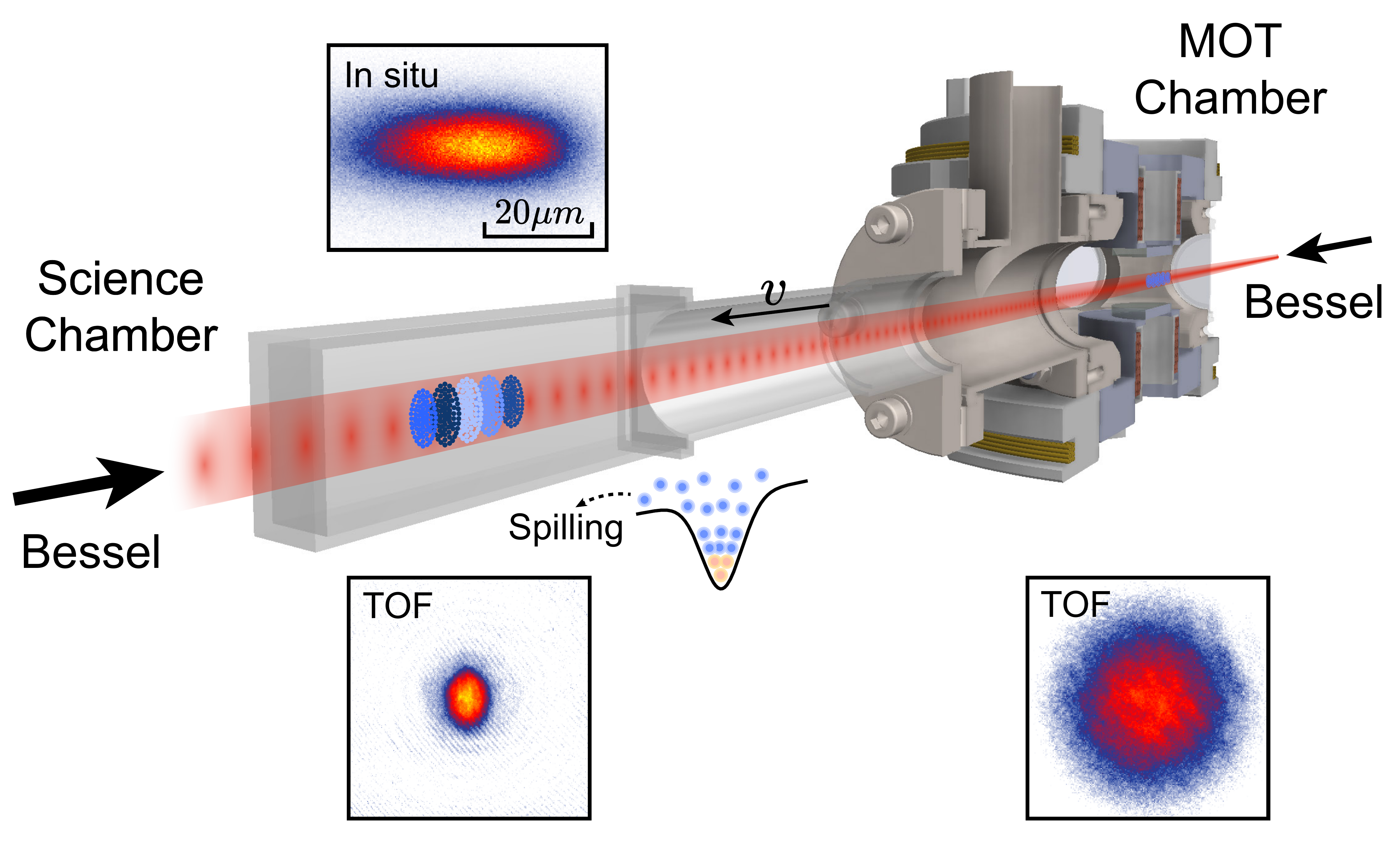}
    \caption{\textbf{Experimental setup}. Cold ytterbium atoms are transported from a magneto-optical trap (MOT) chamber to a science chamber over a distance of 34$\,$cm within 350$\,$ms by a moving optical lattice formed by two counter-propagating Bessel beams. The lattice moves at a velocity of $v=\lambda \Delta f/2$ where $\lambda$ is the lattice wavelength and $\Delta f$ is the frequency difference between the two beams. The transport is precisely controlled by tuning the frequency difference. In the final stage of transport, the trap potential is tilted via deceleration and reduced by lowering the power, allowing hotter atoms to spill out and thereby evaporatively cooling the remaining atoms. Time-of-flight (TOF) images are taken after 15$\,$ms and 20$\,$ms of expansion in the MOT after loading and science chambers after phase synchronization, respectively.}
    \label{fig:Experimental setup}
\end{figure}

Fig.~\ref{fig:Experimental setup} shows the setup where cold atoms are transported over $34\,$cm between the magneto-optical trap (MOT) chamber and the science chamber which is a glass cell. Cold atoms are first prepared by a two-color MOT~\cite{li2025two}, followed by a compressed MOT to load atoms into a dipole trap, and short forced evaporative cooling to prepare different initial temperatures. Then, atoms are adiabatically loaded into the one-dimensional moving optical lattice by simultaneously ramping up the two counter-propagating Bessel beams to a lattice depth of $U/k_{\text B}\approx15\,\mu$K with the Boltzmann constant $k_{\text B}$.

A key requirement for fast cold atom transport is that the moving optical lattice provides sufficient and uniform potential along the transport, in particular for cold atoms near quantum degeneracy which remains highly challenging due to sensitivity to vibrational noise and trap nonuniformity along the transport direction. For instance, magnetic transport relying on a series of coils typically restricts optical access, and technical noise and mechanical vibrations limit the achievable cold temperature~\cite{hansel2001magnetic}. On the other hand, transport based on tunable lenses typically has weak axial confinement and vibrational noise~\cite{leonard2014optical}, while a moving optical lattice formed by Gaussian beams inevitably diverges over long transport distances and results in nonuniform trap potential along the transport~\cite{schrader2001applied,schmid2006long,klostermann2022fast,bao2022fast,trisnadi2022design,matthies2024long}.

To overcome the excessive diffraction of Gaussian beams over long distances, we employ Bessel beams which maintain a relatively uniform, diffraction-free trapping potential along the entire transport. The Bessel beam is generated by axicons~\cite{brzobohaty2008high} and the diffraction-free range is $z_{\max} = w_0 / \tan(\beta)$, where $w_0$ is the incident Gaussian waist and $\beta = (n-1)\alpha$ with axicon apex angle $\alpha$ and refractive index $n$. The angle of axicon used in the experiment is 1$^\circ$.
The wavelength of the moving lattice is 1064$\,$nm, far detuned from the $^1S_0 \to {}^1P_1$ transition at 399$\,$nm and the $^1S_0 \to {}^3P_1$ transition at 556$\,$nm of Yb atoms, thereby minimizing photon scattering and suppressing heating induced by the moving lattice.
Two incident beams have powers of 14$\,$W and 17$\,$W, with beam waists of 2.5$\,$mm and 3$\,$mm, respectively. The resulting Bessel beams have central spot radii of $\sim50\,\mu$m, providing strong axial gradients against gravity. Fig.~\ref{fig:Transport_Control}(a) in the Supplemental Material shows the details of the corresponding trap potential of the two individual beams, the total trap depth and lattice depth. In addition to its ability to handle higher optical power, this scheme is more power-efficient than Gaussian-beam transport. For the same transport distance of $34\,$cm, a Gaussian beam would require a beam waist of $240\,\mu$m to achieve a Rayleigh range of $17\,$cm, thereby significantly increasing the required optical power for trapping atoms.

The alignment of the moving lattice requires stringent procedures as the central waists of both Bessel beams are only 50$\,\mu$m. We developed a systematic procedure to align the moving lattice using multiple imaging techniques and lattice diffraction effect (see details in the Supplemental Material). Both the radial and lattice-induced axial confinement maintain the atomic cloud dense. The inset of Fig.~\ref{fig:Experimental setup} shows the atomic cloud size at the end of transport detected by high-resolution insitu imaging. 

The velocity of the moving lattice, $v=\lambda \Delta f/2$ is controlled by the frequency difference between two Bessel beams, where $\Delta f$ is set by double-pass acousto-optic modulators (AOMs) and $\lambda$ is the lattice wavelength. Transport of cold atoms begins with constant acceleration, followed by constant velocity transport and deceleration into the science chamber (see Supplemental Material for the detailed sequence). Since the axicons are purely static optical elements, our scheme has less vibration and technical noise than mechanical approaches using translation stages, rotating Moire lenses, or liquid tunable lenses. Our scheme provides precise positional control of the atoms. By comparing the designed motion curve with the measured center-of-mass of atoms, we achieved a correlation coefficient of $r^2 > 0.9999$ with corresponding maximum position residual less than half the camera pixel size. The actual transport motion precision is around 2 $\mu$m over long distance transport (see details in the Supplemental Material). After transport, atoms are loaded into a crossed dipole trap in the science chamber, where they are held for phase synchronization.

\section{Transport and Evaporation}

\begin{figure}[htbp]  
    \centering
    \includegraphics[width=1\linewidth]{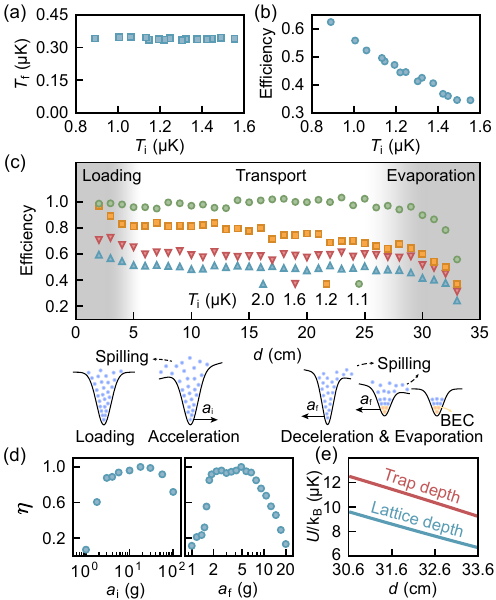}
    \caption{\textbf{Transport and cooling mechanism}. \textbf{(a,b)} The final temperature and transport efficiency for different initial loading temperatures. The overall transport efficiency increases with decreasing initial temperature. The final temperature depends on the final trap depth rather than the initial loading. \textbf{(c)} The round-trip transport efficiency as a function of transport distance for different initial temperatures. The schematic illustrates the mechanisms of atom loss and cooling during the loading phase as well as the effective evaporative cooling in the final stage. \textbf{(d)} The accelerations and final decelerations in units of the gravitational acceleration $g$ are optimized based on the relative atom number after transport. \textbf{(e)} The trap depth and optical lattice depth at different positions in the final cooling stage.}
    \label{fig:Transport Mechanism}
\end{figure}

Fig.~\ref{fig:Transport Mechanism} shows the calibration of the transport efficiency. Transport of cold atoms consists of three stages, initial loading, transport, and final unloading. Notably, when the lattice is accelerated or decelerated, the trapping potential of each pancake is tilted, effectively reducing the lattice depth while the trap frequencies remain high, a condition favorable for evaporative cooling. A similar method has been reported by evaporative cooling using magnetic gradient potential~\cite{hung2008runaway}.

The initial loading temperature is controlled by varying the evaporation endpoint in the dipole trap. We characterize the temperature and efficiency of the transported atoms as functions of initial temperature in Fig.~\ref{fig:Transport Mechanism}(a) and (b). The final temperature is less dependent on the initial temperature and approximately 340$\,$nK. We further calibrate the overall transport efficiency at different positions by performing round-trip transport of the atoms, i.e. moving atoms back and forth over different distances in Fig.~\ref{fig:Transport Mechanism}(c). Even after short round trips, atomic clouds with different initial conditions reach similar final temperatures, whereas hotter atoms are lost significantly, as reflected in the reduced transport efficiency. This initial atom loss originates from the population of higher lattice bands for hot temperatures. Subsequent acceleration effectively tilts the lattice potential, causing the hotter atoms to spill out. In the left panel of Fig.~\ref{fig:Transport Mechanism}(d), we optimized the relative final atom number by varying the initial acceleration. For small accelerations, the atomic cloud suffers severe losses due to three-body interactions, while for larger accelerations, over-tilting of the lattice and heating also lead to decreased atom numbers.

The second stage spans the transport of approximately 25 cm indicated by the white regime in Fig.~\ref{fig:Transport Mechanism}(c). Both the atom number and temperature remain stable and exhibit a slight decrease during round-trip transport, with the transport efficiency approaching 90\%. Atoms with different initial temperatures have almost identical return temperatures, indicating that the temperature during transport is mainly determined by the lattice depth. The dependence of the round-trip temperature on the transport distance is shown in Fig.~\ref{fig:Temperature} in the Supplemental Material.

The final evaporative cooling occurs in the last few centimeters of the transport path. The trapping potential is below that of the initial loading stage and decreases substantially with distance (see Supplemental Material for details). The atoms are decelerated by reducing the frequency difference $\Delta f$. The final temperature is mainly determined by the lattice depth at the destination. Total transport efficiency decreases with increasing initial temperature, indicating that hotter atoms are removed during the transport. Fig.~\ref{fig:Transport Mechanism}(e) shows both the trap depth and the lattice depth of the transport lattice decrease, facilitating thermalization and evaporative cooling through the removal of hotter atoms. The right panel of Fig.~\ref{fig:Transport Mechanism}(d) shows the dependence of the relative final atom number on the deceleration. Here, the deceleration is optimized around $a_f\approx5g$ in the experiment achieving more effective cooling of atoms.

This evaporative cooling and the strong axial confinement of the optical lattice due to the small beam size of the Bessel beams, result in a sufficient phase-space density for subsequent collision thermalization. We managed to transport $3\times10^5$ atoms at 340$\,$nK to the destination. Due to strong confinement along the transport direction, there are approximately 57 isolated, pancake-shaped two-dimensional clouds detected by the insitu image with random phases as a result of independent collision thermalization during evaporation.

Notably, atoms in each pancake cannot undergo Bose-Einstein condensation due to the quasi-2D geometry ($\hbar\omega_z/(k_{\text B} T)\approx 3.5$). Atoms have a similar phase within each pancake but different phases between different pancakes. The temperature of atoms $T=340~\mathrm{nK}$ is in between the quasicondensation crossover temperature, $\tilde T_c=1.4~\mathrm{\mu K}$ and the Berezinskii-Kosterlitz-Thouless (BKT) transition temperature $T_{\mathrm{BKT}}=201~\mathrm{nK}$ ~\cite{BEC&SF,fisher1988dilute,clade2009observation}; see Supplemental Material for details. The quasicondensate fraction in our experiment is $n_{2s}^{(0)}/n=0.29$, which is further confirmed by the emergence of a narrow peak in 10$\,$ms TOF images.

\begin{figure}[htbp]
    \centering
    \includegraphics[width=\linewidth]{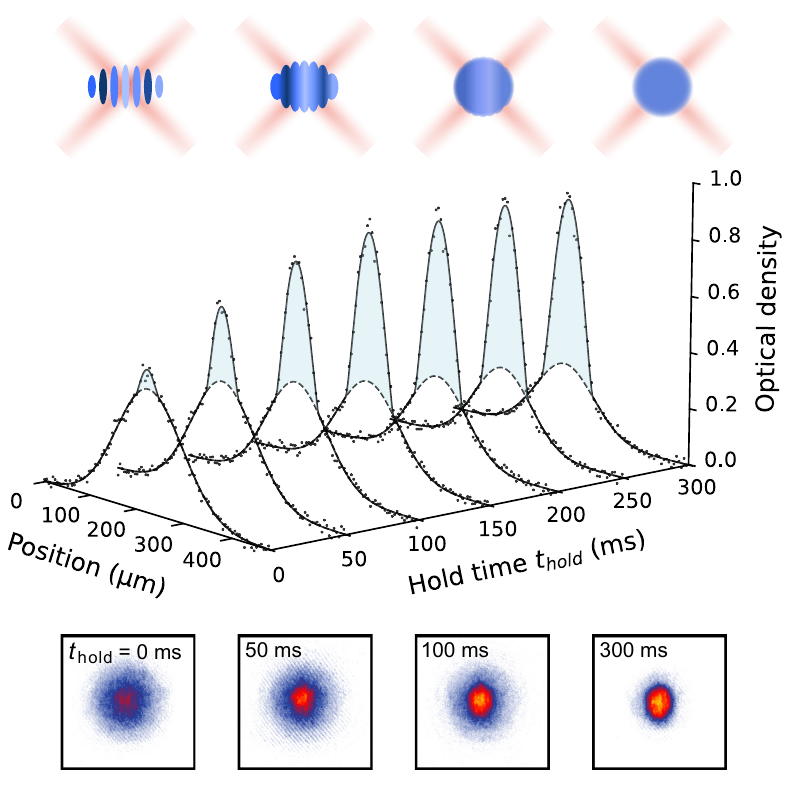}
    \caption{\textbf{Emergence of BEC from the phase synchronization.} Cold atoms with initially different phases (indicated by different blue colors) are released from the moving lattice to a dipole trap and subsequently synchronized through atomic interactions, resulting in the formation of a Bose-Einstein condensate. Bimodal fits (solid lines) to the momentum distributions at different hold times, with the Gaussian component (dashed lines) representing the thermal atoms, are used to extract the condensate and thermal fractions. The TOF is 20$\,$ms.} 
    \label{fig:Transport BEC}
\end{figure}

\section{Emergence of BEC from Phase Synchronization}

Eventually, transferred atoms in each pancake are further loaded from the lattice into a crossed 1064$\,$nm dipole trap in the science chamber with trap frequencies $(\omega_x,\omega_y,\omega_z)=2\pi\times(121, 73, 142)$$\,$Hz, and potential depth $U_{\text{sci}}\approx5.5\,\mu$K. Then, atoms are held for variable times, and a BEC emerges from the phase synchronization among the initial quasi-2D pancakes. Similar to the Kibble-Zurek mechanism, phase domains formed in the moving lattice subsequently relax and synchronize through atomic interactions. These interactions help thermalization, synchronize the phases across all pancakes, induce spontaneous breaking of the U(1) symmetry, and eventually lead to the formation of a Bose-Einstein condensate.

\begin{figure}[htbp]
    \centering
    \includegraphics[width=\linewidth]{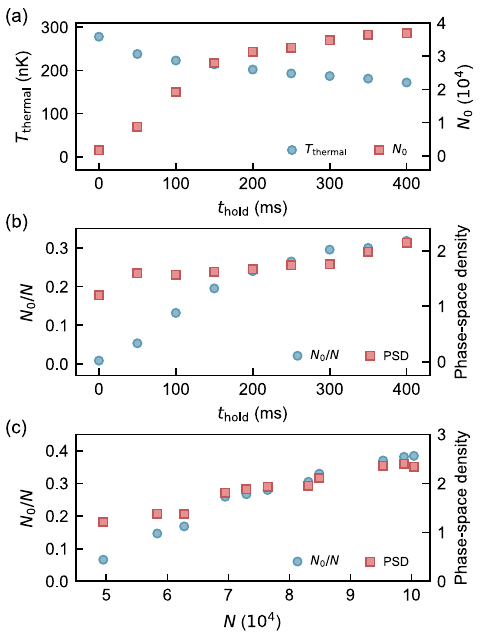}
    \caption{\textbf{Benchmark of BEC formation} \textbf{(a)} The temperature of the thermal cloud decreases and the condensate fraction increases, with increasing hold time. \textbf{(b)} Corresponding condensate fraction and phase-space density (PSD) extracted from fittings are plotted as a function of hold time. \textbf{(c)} The condensate fraction and PSD as a function of the total atom number remaining after the $0.3\,$s hold. Increasing the atom number loaded into the moving lattice leads to a higher final condensate fraction and a larger PSD after synchronization.}
    \label{fig:BEC benchmark}
\end{figure}

Fig.~\ref{fig:Transport BEC} shows bimodal fits to momentum distributions together with representative TOF images after different hold times, revealing the rapid formation of a distinct condensate peak (light blue) from the thermal Gaussian background. This indicates BEC growth through collision thermalization in the dipole trap, with the condensate developing rapidly within 150$\,$ms and gradually approaching saturation thereafter.

We calibrate the temperature and condensation throughout the phase synchronization process in Fig.~\ref{fig:BEC benchmark}. The dependence of temperature $T_{\text{thermal}}$ and condensed atom number $N_0$ on the hold time shows a different mechanism from conventional evaporative cooling. During the first 150$\,$ms, the temperature decreases slowly but the condensed atom number increases rapidly and saturates at approximately $4\times10^{4}$. The final temperature is $T=172\,$nK and the critical temperature for the BEC transition is $T_c=209\,$nK. Fig.~\ref{fig:BEC benchmark}(b) presents the condensate fraction and the phase-space density (PSD) as a function of hold time, $\text{PSD}=\ N_{\text{thermal}}\left(\hbar\bar\omega/k_BT\right)^3$ with $\bar\omega=(\omega_x\omega_y\omega_z)^{1/3}$ and the number of thermal atoms $N_{\text{thermal}}$. Interestingly, the PSD increases slowly throughout the process, whereas the BEC fraction grows sharply. Moreover, both the condensate fraction and PSD increase with increasing the total number of transported atoms. The condensate fraction is measured as a function of the total atom number remaining after the $0.3\,$s hold in Fig.~\ref{fig:BEC benchmark}(c). With increasing the atom number, both the condensation and PSD increase.

\section{Conclusion}

We demonstrate rapid optical transport of cold atoms to quantum degeneracy using an optical moving lattice formed by interfering two Bessel beams. A gas of $3\times10^5$ Yb atoms at 340$\,$nK is transferred over 34$\,$cm in 350$\,$ms with above $60\%$ efficiency and well controlled by tuning the frequency difference between two beams. Effective evaporative cooling of atoms in each pancake is achieved by decelerating the lattice while gradually lowering the trapping potential along the transport path. A gas of $1\times10^5$ atoms with $40\%$ condensate eventually emerges from the phase synchronization across those pancakes with different phases. This phase synchronization process opens new opportunities to study BKT and BEC transitions. Our rapid and efficient transport demonstrated here paves the way for atom lasers and continuously operated large-scale atomic arrays.

\section{Acknowledgments} 

This work was supported by the Quantum Science and Technology-National Science and Technology Major Project (2024ZD0301800), the Beijing Natural Science Foundation (Z240007), the National Natural Science Foundation of China (12374242), and the Scientific Research Innovation Capability Support Project for Young Faculty (ZY2025014).

\section{Data Availability}

The data presented in this work are available on the Peking University Open Research Data Platform.

\bibliography{OT}

\section*{Supplemental Material}
\setcounter{equation}{0}
\setcounter{figure}{0}
\setcounter{table}{0}
\setcounter{section}{0}
\renewcommand{\thefigure}{S\arabic{figure}}
\renewcommand{\theequation}{S\arabic{equation}}
\renewcommand{\theHfigure}{S\thefigure}

\section{Bessel Beams and Alignment}
\begin{figure}[htbp]
    \centering
    \includegraphics[width=1\linewidth]{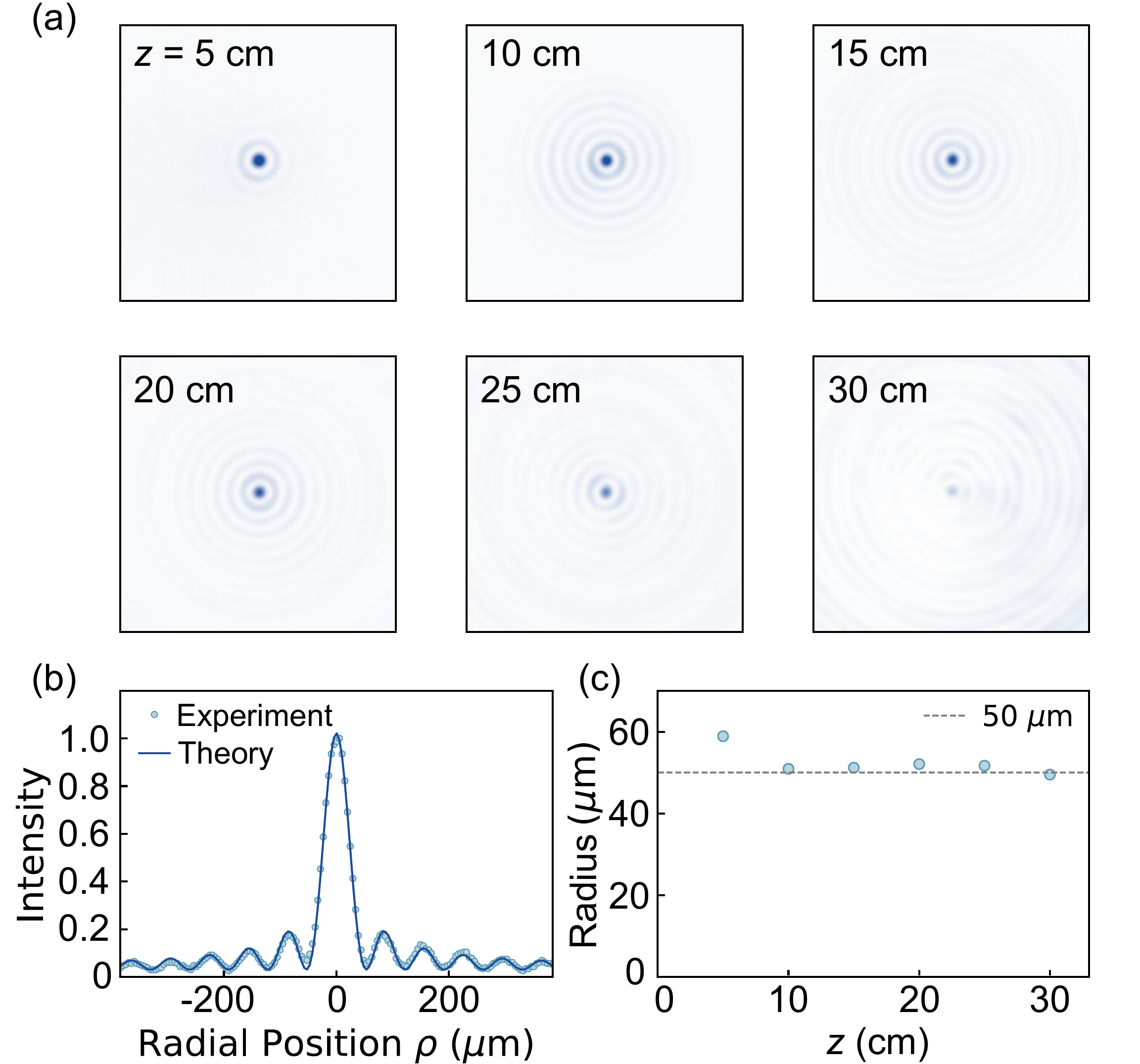}
    \caption{\textbf{Calibration of the Bessel beam}. (a) The measured beam profile of a collimated Gaussian beam with a waist radius of 2.5$\,$mm after transmission through the axicon with an apex angle of $\alpha = 1^{\circ}$ and a refractive index of $n=1.5$. (b) The measured intensity (circles) agrees well with the simulation (solid lines). (c) The beam waist of the Bessel beam is almost constant $\sim50\,\mu$m along with transport.}
    \label{fig:bessel}
\end{figure}

The moving lattice consists of two Bessel beams. The Bessel beams are generated by passing collimated Gaussian beams through axicons. Within a distance $z_{\max}=w_0/\tan \beta$ after the axicon, the central spot size of the Bessel beam remains almost the same, where $w_0$ is the incident Gaussian beam waist radius and $\beta=(n-1)\alpha$ with the axicon apex angle $\alpha$ and the refractive index $n$. The intensity of the Bessel beam as a function of the radial coordinate $\rho$ and the axial coordinate (along the transport direction) $z$ is given by~\cite{brzobohaty2008high}
\begin{equation}
I(\rho, z) = \frac{4 P k \sin \beta}{w} \frac{z}{z_{\max }} J_{0}^{2}(k \rho \sin \beta) \exp \left(-\frac{2 z^{2}}{z_{\max }^{2}}\right)
\end{equation}
where $J_0$ is the zeroth-order Bessel function of the first kind, $P$ is the power and $k = 2\pi / \lambda$ is the wavevector with the wavelength of the lattice beam $\lambda$. The zero-order Bessel spot corresponds to the diffraction-free central lobe, with a radius $\rho_0 \approx 2.405 / k\sin \beta=50\,\mu$m in our experiment. We detected the profiles of the Bessel beams at different positions after the axicon in Fig.~\ref{fig:bessel}(a). We fit the Bessel curve to the measurement and extract the beam size as a function of distance from the axicon $z$ in Fig.~\ref{fig:bessel}(b,c). The measurement and theory are in good agreement.

\begin{figure}[htbp]
    \centering
    \includegraphics[width=1\linewidth]{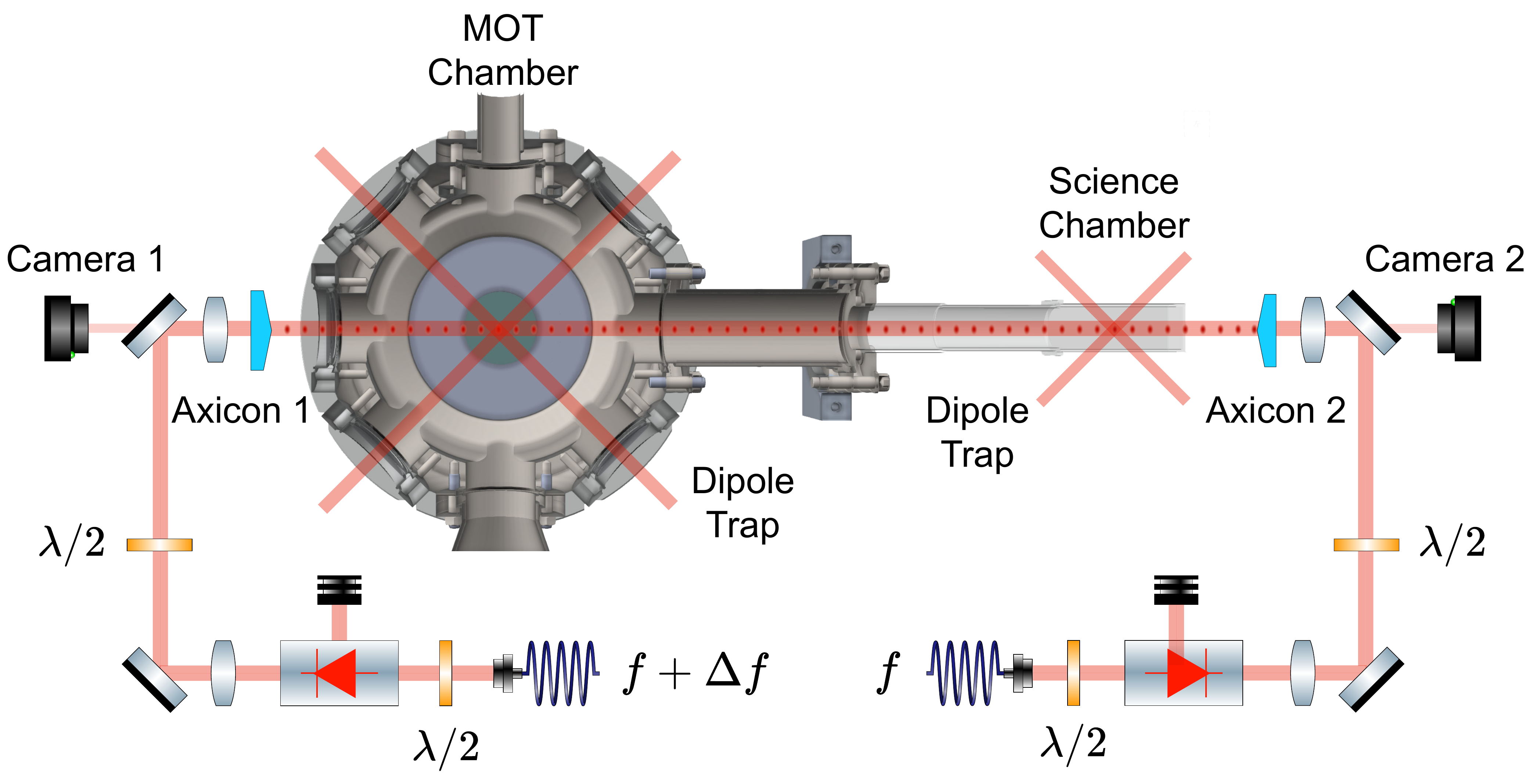}
    \caption{\textbf{Experimental details and alignment procedure of two Bessel beams}. The Bessel beams are generated from Gaussian beams delivered by fiber couplers and optical isolators, and subsequently expanded to beam waist radii of 2.5 mm and 3 mm in the left and right optical arms, respectively, with the beam profiles monitored by cameras via beam splitters. }
    \label{fig:Precise Alignment}
\end{figure}

As it is challenging to align two beams passing through two holes with a radius of $50\,\mu$m over $34\,$cm distance, we develop a systematic and efficient method for precise long-distance alignment of two counter-propagating Bessel beams, as illustrated in Fig.~\ref{fig:Precise Alignment}. First, we align the two counter-propagating Gaussian beams to completely overlap in the absence of axicons, and ensure that each beam fully enters the opposite optical isolator. Next, we insert Axicon~1, and simultaneously adjust the preceding pair of mirrors and the axicon itself to produce a uniform Bessel beam capable of independently trapping cold atoms in the crossed dipole trap in the MOT chamber while still entering the isolator. This step relies on different imaging systems in our optical setup, iteratively comparing the positions of the cold atoms in the dipole trap, followed by further optimization of the loading efficiency into the Bessel beam.

Subsequently, we use two cameras behind the mirrors to record the Bessel ring patterns formed by each counter-propagating beam after passing through Axicon~1. Multiple concentric ring images are fitted to determine their central positions. After adding Axicon~2, we ensure that the Bessel beam center on Camera~2 remains unchanged with and without the axicon, guaranteeing that Axicon~2's center aligns with the central spot of Bessel~1 within $\sim20\,\mu$m. Finally, we adjust the pair of mirrors after Axicon~2 to achieve uniform imaging of the retro-reflected Gaussian~2 beam (passing through both axicons) on Camera~1. 
Consequently, Bessel~1 is centered through both axicons relative to the atomic position. Empirically, this coarse alignment already enables atomic transport over distances of up to approximately 15 cm.

Finally, fine alignment is achieved by adjusting the pair of mirrors after Axicon~2 until the cold atoms in the MOT chamber exhibit a pronounced diffraction signal, i.e., the Kapitza-Dirac (KD) effect, induced by the moving lattice. This procedure ensures precise alignment in the MOT chamber. At this stage, atoms can be reliably transported to the science chamber. Further optimization of the transport efficiency, together with fine adjustments of Axicon~2 and the subsequent pair of mirrors, ultimately achieves precise alignment of the two Bessel beams.

\section{Transport Equations}
Here we provide the detailed calculation of the moving optical lattice. The complex amplitudes of the two beams are given by

\begin{equation}
\begin{aligned}
E_1(z, \rho, t) &= E_1(z,\rho)  e^{i(\omega + \Delta \omega) (t - z/c)}, \\
E_2(z, \rho, t) &= E_2(z,\rho) e^{i\omega(t + z/c)}, \\
E_{\text{total}}(z, \rho, t)&=E_1(z, \rho, t)+E_2(z, \rho, t),
\end{aligned}
\end{equation}
where 
$\omega$ is the optical angular frequency of beam 2, and $\Delta\omega$ is the angular frequency difference between two beams. The intensity distributions of the two Bessel beams read

\begin{equation}
\begin{split}
I_1(z, \rho) &= E_1(z, \rho, t)E_1^*(z, \rho, t) \\
&= \frac{4 P_1 k \sin \beta}{w_{1}} \frac{z}{z_{\max1}} J_{0}^{2}(k \rho \sin \beta) \exp \left(-\frac{2 z^{2}}{z_{\max1}^{2}}\right), \\
\\
I_2(z, \rho) &= E_2(z, \rho, t)E_2^*(z, \rho, t) \\
&= \frac{4 P_2 k \sin \beta}{w_{2}} \frac{l-z}{z_{\max2}} J_{0}^{2}(k \rho \sin \beta) \exp \left(-\frac{2 (l-z)^{2}}{z_{\max2}^{2}}\right),
\end{split}
\end{equation}
where $P_i$ and $w_i$ are the power and waist of the incoming Gaussian beams after each axicon, $z_{\max i}=w_i/ \tan(\beta)$ is the diffraction-free propagation range, and $l$ is the distance between the two axicons.

The interference of the two Bessel beams produces the intensity distribution given by $I(z, \rho, t) = E_{\text{total}}(z, \rho, t) E_{\text{total}}^*(z, \rho, t)$. The resultant intensity is the sum of the individual beam intensities and an interference term,
         
\begin{equation}
\begin{aligned}
I(z, \rho, t) = I_1(z, \rho) + I_2(z, \rho) \\
+ 2\sqrt{I_1(z, \rho)I_2(z, \rho)} \cos(\Delta\phi),
\end{aligned}
\end{equation}
where the phase difference is $\Delta\phi = \Delta\omega t - 2k z$. Therefore, the lattice transport velocity is as follows,

\begin{equation}
v = \frac{1}{2}\lambda \cdot \Delta f
\end{equation}

The dipole potential of the moving lattice is proportional to the light intensity,
\begin{equation}
U_{\text{pot}}(z, \rho, t) = -\frac{3\pi c^2}{2\omega_0^3} \left( \frac{\Gamma}{\Delta} \right) I(z, \rho, t)=-\alpha I(z, \rho, t)
\end{equation}
where $\alpha$ is the atomic polarizability. In the axial direction along the transport direction, the lattice potential is sinusoidal and its depth is
\begin{equation}
U_{\text{axial}}(z)= -4\alpha \sqrt{I_1(z)I_2(z)}.
\end{equation}

For the radial direction, perpendicular to the transport direction, the optical potential traps the atoms to compensate the gravity. The potential depth is determined by the total light intensity,

\begin{equation}
U_{\text{radial}}(z) = -\alpha \left[I_1(z) + I_2(z) + 2\sqrt{I_1(z)I_2(z)}\right].
\end{equation}

Fig. \ref{fig:Transport_Control}(a) shows calculated axial lattice and the radial trapping potential over the whole transport. 

\begin{figure}[htbp]
    \centering
    \includegraphics[width=\linewidth]{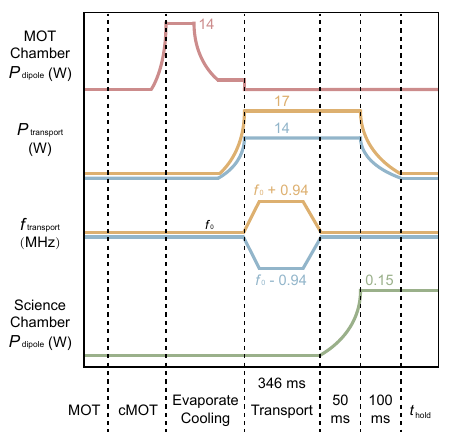}
    \caption{\textbf{Experimental sequence.} The powers and frequencies of the moving lattice, as well as the powers of the dipole traps in the MOT and science chambers, are shown for the different stages of the experiment.} 
    \label{fig:Sequence}
\end{figure}

\section{Experimental Sequence and Precise Transport Control}

Fig.~\ref{fig:Sequence} shows the experimental sequence. A gas of $1 \times 10^8$ $^{174}$Yb atoms at 18~$\mu$K is prepared by Zeeman slowing and two-color magneto-optical trapping~\cite{li2025two}. The atoms are further compressed via a compressed MOT stage (cMOT) and transferred into a crossed 1064$\,$nm optical dipole trap, with each beam having a power of 14$\,$W. The dipole trap is exponentially ramped up in 10$\,$ms and the temperature of atoms is controlled by the endpoint of evaporative cooling in the dipole trap. Subsequently, the moving lattice is ramped adiabatically to transfer atoms from the dipole trap into the moving optical lattice.

Next, the frequencies of the two counter-propagating transport beams are tuned in time to move the optical lattice and transport the atoms to the science chamber within $350\,$ms.
The transport frequencies are tuned by double-pass AOMs and the two beams are symmetrically detuned by $\pm0.94\,$MHz (Fig.~\ref{fig:Sequence}).
The total frequency difference between two beams $\Delta f=1.88$\,$\mathrm{MHz}$.
This sets the lattice velocity $v=\lambda\Delta f/2=1.0\,$m/s.

The two dipole beams are ramped up to 150$\,$mW within 50$\,$ms, followed by ramping down the two transport beams in 100$\,$ms. Finally for the stage of phase synchronization, atoms are held in the dipole trap for different hold times \( t_{\text{hold}} \).

\begin{figure}[htbp]
    \centering
    \includegraphics[width=\linewidth]{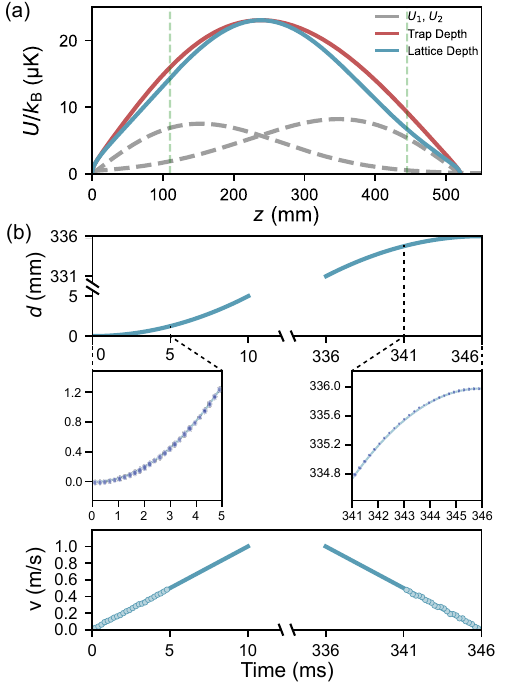}
    \caption{\textbf{Precise control of cold atom motion} \textbf{(a)} Trap depth (red) and lattice depth (blue) of the moving lattice. $U_1$ and $U_2$ denote the potential of two individual Bessel beams. The green dashed lines indicate the positions of the MOT chamber (left) and the science chamber (right). \textbf{(b)} Measured position (insets) and extracted velocity (circles) of transferred atoms at different time agree well with the designed curves (solid lines). Data for acceleration and deceleration are probed in MOT and science chambers respectively.}
    \label{fig:Transport_Control}
\end{figure}

Crucially, the transport of cold atoms is well controlled by the frequency difference between two Bessel beams, including 10$\,$ms linear acceleration to 1$\,$m/s, 326$\,$ms constant-velocity transport, and 10$\,$ms deceleration to rest in Fig.~\ref{fig:Transport_Control}b. The atomic position during acceleration and deceleration phases is detected in the MOT and science chambers respectively. The measured center of mass in the acceleration and deceleration stages agrees well with the designed trajectories with correlation coefficients $r^2>0.9999$ for both stages. We estimate that the actual accuracy of transport is about $2\,\mu\mathrm{m}$ over the whole transport.

\section{Round-trip transport}

\begin{figure}[htbp]
    \centering
    \includegraphics[width=\linewidth]{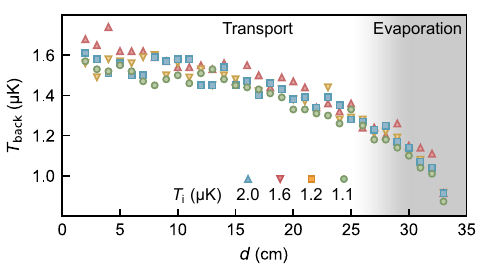}
    \caption{\textbf{Temperature after round-trip transport.} The final temperature is independent of the initial loading temperature and is determined by the trap potential.}
    \label{fig:Temperature}
\end{figure}

Fig.~\ref{fig:Temperature} shows the temperature of atomic clouds for different initial temperatures after round-trip transport to different positions. Similar to Fig.~\ref{fig:Transport Mechanism}(c) in the main text, the transport and evaporation stages are separated. During the second phase of transport, the temperature gradually decreases, while in the final phase, it drops significantly, indicating a collision thermalization in the evaporative cooling. Moreover, for atoms with different initial temperatures, the round-trip temperatures are almost the same, indicating that the temperature is governed by the local trap depth.

\section{Thermometry and Condensation}

The strong lattice confinement along the moving lattice results in a series of quasi-2D pancakes, following the 2D criterion $\hbar\omega_z/(k_{\text B} T)\approx 3.5$, where $\omega_z$ is the axial trap angular frequency, $\hbar$ is the reduced Planck constant, $k_{\text B}$ is the Boltzmann constant, and the temperature after transport is $T=340\,$nK. In this regime, the coupling between pancakes is frozen and the isolated 2D pancakes cannot have true long-range order. To quantify phase coherence and quasicondensation within each pancake, we consider two characteristic temperatures: The quasicondensation crossover temperature $\tilde T_c$, and the Berezinskii-Kosterlitz-Thouless (BKT) transition temperature $T_{\mathrm{BKT}}$~\cite{BEC&SF,fisher1988dilute,clade2009observation}.

Based on the Landau quasiparticle theory for vanishing 2D superfluid density $n_{2s}$, the critical temperature from a thermal gas to a quasicondensate gas is
\begin{equation}
\tilde T_c=\frac{2\pi\hbar^2 n_2}{k_{\mathrm{B}} m}
\left[\ln\!\left(\frac{\hbar^2 n_2}{m^2c^2}\right)\right]^{-1},
\end{equation}
where $c$ is the sound speed of interacting atoms in two dimensions, $m$ is the atomic mass and $n_2$ is the peak 2D number density in each pancake. We calculate $c$ from the equation of state using $m c^2\approx \mu\approx g_{\mathrm{2D}} n_2$, where $\mu$ is the chemical potential and $g_{\mathrm{2D}}$ is the effective 2D coupling constant.

The BKT transition temperature is
\begin{equation}
T_{\mathrm{BKT}}=\frac{2\pi\hbar^2 n_2}{k_{\mathrm{B}} m}
\left[\ln\!\left(\frac{C\hbar^2}{m g_{\mathrm{2D}}}\right)\right]^{-1},
\end{equation}
with $C\approx380$~\cite{prokof2001critical}. In our experiment, the final gas contains $N=3\times10^5$ atoms distributed across $57$ pancakes, yielding an average atom number per pancake $N_{\mathrm{p}}= 5.3\times10^3$. We estimate $\tilde T_c= 1.4~\mathrm{\mu K}$ and $T/\tilde T_c= 0.24$, while $T_{\mathrm{BKT}}= 201~\mathrm{nK}$ and $T/T_{\mathrm{BKT}}= 1.7$. Thus the temperature of atoms in each pancake is close to the BKT transition but below the quasicondensation crossover temperature.

\begin{figure}[htbp]
    \centering
    \includegraphics[width=1\linewidth]{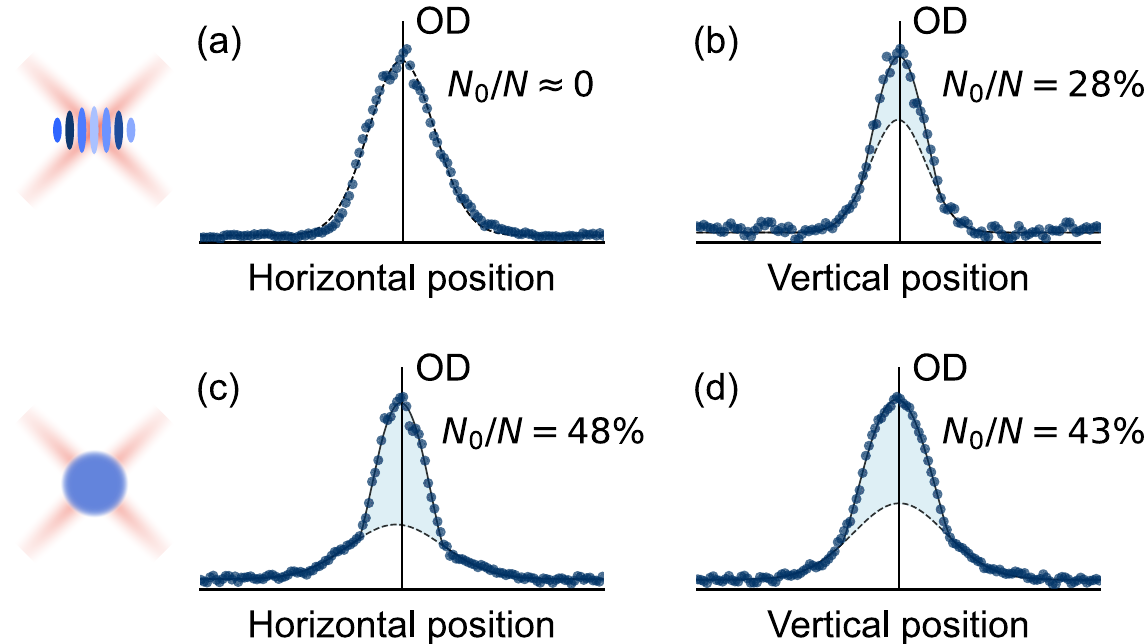}
    \caption{\textbf{Momentum distributions of atoms in the pancake after transport and in the dipole trap after phase synchronization.} \textbf{(a,b)} Cuts along the transport direction (horizontal) and perpendicular to the transport direction (vertical) of the momentum distribution of atoms in pancakes after transport. In contrast to the vertical momentum distribution which shows a clear bimodal feature (solid lines) and quasicondensation (light blue), the horizontal momentum distribution remains thermal and is well fitted by a Gaussian profile (dashed lines). The extracted quasicondensate fraction from the bimodal fitting (solid lines) is $0.28$, in good agreement with the estimate $n_{2s}^{(0)}/n_2=0.29$. \textbf{(c,d)} After phase synchronization, both the horizontal and vertical distributions exhibit clear bimodal features with condensate fractions of $N_0/N=48\%$ and $43\%$, in good agreement with the estimate $N_0/N=46\%$. TOF are 10ms and 20ms for (a,b) and (c,d) respectively.}
    \label{fig:Quasicondensation}
\end{figure}

Correspondingly, we estimate the quasicondensate fraction in each pancake, i.e. the ``bare'' superfluid fraction, $n_{2s}^{(0)}/n_2=0.29$ at $T=340~\mathrm{nK}$, where the superscript $(0)$ denotes the Landau estimate without considering vortices. We further confirm it by observing a condensed momentum peak after a short TOF along the direction perpendicular to the lattice in Fig.~\ref{fig:Quasicondensation}.

After transferring atoms into the shallow crossed dipole trap, atomic interactions help synchronize phases across pancakes. During this collision thermalization, rapid phase synchronization leads to the emergence of a global BEC on a timescale of $100\,\mathrm{ms}$. The critical temperature for the BEC transition in the harmonic trap is
\begin{equation}
T_c=\frac{\hbar\bar\omega}{k_{\mathrm{B}}}\left(\frac{N_{}}{\zeta(3)}\right)^{1/3},
\end{equation}
where $\bar\omega=(\omega_x \omega_y \omega_z)^{1/3}$ is the average trap frequency and $\zeta$ is Riemann zeta function. The critical temperature for BEC transition in the final dipole trap is $T_c=209\,\mathrm{nK}$. We achieve a gas of $N=10^5$ atoms at a temperature of $T=172\,$nK after phase synchronization. The measured condensate fraction is approximately $40\%$, in agreement with the estimate $N_0/N=1-(T/T_c)^3=46\%$ based on temperature.

Fig.~\ref{fig:Quasicondensation} shows the thermometry of atoms in the pancake and in the dipole trap after phase synchronization. For the pancake, there is a clear difference between the cut of $10\,\mathrm{ms}$ TOF along the transport direction (horizontal) and the direction perpendicular to transport (vertical). In contrast to the vertical direction, the quasicondensate fraction of the horizontal direction is almost zero. We fit the bimodal curve (solid lines) to the data to extract the quasicondensate fraction $0.28$, consistent with the simulated value $n_{2s}^{(0)}/n_2=0.29$. For the atoms in the dipole trap after a hold of $0.3\,\mathrm{s}$, condensation emerges in both horizontal and vertical directions with a condensate fraction of $48\%$ and $43\%$ respectively in the $20\,\mathrm{ms}$ TOF image in Fig.~\ref{fig:Quasicondensation}(c,d).

\end{document}